%% file: ms.tex
\documentclass[prd,11pt]{article}
\pdfoutput=1 

\usepackage{geometry}
\usepackage{authblk}
\usepackage{latexsym}

\usepackage{enumerate}
\usepackage[inline]{enumitem}
\usepackage{amsfonts,dsfont}
\usepackage{nicefrac}
\usepackage{microtype}
\usepackage{mathtools}

\usepackage{sidecap}
\usepackage{caption}
\usepackage{subcaption}
\usepackage{wrapfig}
\usepackage{enumitem}
\usepackage[linesnumbered,ruled,noline,noend]{algorithm2e}
\usepackage{amsmath,amssymb,graphicx,multirow,xspace,slashed}
\usepackage[colorlinks=true,urlcolor=blue,anchorcolor=blue,citecolor=blue,filecolor=blue,linkcolor=blue,menucolor=blue,pagecolor=blue]{hyperref}
\usepackage[normalem]{ulem}
\usepackage{amssymb}
\usepackage{multicol}
\usepackage{adjustbox}
\usepackage{multirow}
\usepackage{color}
\usepackage{xspace}
\usepackage{CJKutf8}

\PassOptionsToPackage{numbers}{natbib}
\usepackage{natbib}
\usepackage[colorinlistoftodo   s,prependcaption,textsize=tiny]{todonotes}

\usepackage{user_definitions}
\usepackage{xspace}
\usepackage[no end]{algpseudocode} 
\usepackage{mathtools}
\usepackage{sidecap}
\usepackage{booktabs}
\algdef{SE}[DOWHILE]{Do}{doWhile}{\algorithmicdo}[1]{\algorithmicwhile\ #1}%

\input{notation.tex}
\def\beq{\begin{equation}}
\def\eeq{\end{equation}}

\usepackage[american]{babel}

\usepackage{natbib} 
\usepackage{mathtools} 
\usepackage{booktabs} 
\usepackage{tikz} 

\title{Reframing Jet Physics with New Computational Methods}
\begin{document}
\vspace{1cm}
\author[1]{Kyle Cranmer}
\author[1]{Matthew Drnevich}
\author[1]{Sebastian Macaluso}
\author[ ]{Duccio Pappadopulo}

\date{ } 
\affil[1]{\small{New York University, USA}}
\providecommand{\keywords}[1]{\textit{Keywords:} #1}
\maketitle

\input{tex/00_abstract}

\vspace{1cm}
\tableofcontents
\vspace{1cm}

\input{tex/01_intro}

\input{tex/02_model_description}

\input{tex/03_greedy_beamsearch}

\input{tex/04_hierarchical-clustering}

\input{tex/05_trellis}
\input{tex/06_ginkgo_RL}

\input{tex/07_a_star}
\input{tex/08_probabilistic_programming}
\input{tex/09_discussion}

\bibliography{references}

\end{document}

%% file: notation.tex
\newcommand{\dataset}{X}

\newcommand{\tree}{\mathtt{H}}
\newcommand{\treeset}{\mathcal{H}}

\newcommand{\EfunS}{\ensuremath{\psi}}
\newcommand{\EfunT}{\ensuremath{\phi}}

\newcommand{\treestar}{\ensuremath{\tree^\star}}

\newcommand{\exactNumClusterings}{(2N-3)!!}

\newcommand{\Ginkgo}{\texttt{Ginkgo}}

\def\beq{\begin{equation}}
\def\eeq{\end{equation}}
\newcommand{\bea}{\begin{eqnarray}\begin{aligned}}
\newcommand{\eea}{\end{aligned}\end{eqnarray}}
\def\bitem{\begin{itemize}}
\def\eitem{\end{itemize}}

\newcommand{\astar}{A*\xspace}

%% file: tex/00_abstract.tex
\begin{abstract}%
We reframe common tasks in jet physics in probabilistic terms, including jet reconstruction, Monte Carlo tuning, matrix element -- parton shower matching for large jet multiplicity, and efficient event generation of jets in complex, signal-like regions of phase space. We also introduce \Ginkgo, a simplified, generative model for jets, that facilitates research into these tasks with techniques from statistics, machine learning, and combinatorial optimization. We review some of the recent research in this direction that has been enabled with \Ginkgo. We show how probabilistic programming can be used to efficiently sample the showering process, how a novel trellis algorithm can be used to efficiently marginalize over the enormous number of clustering histories for the same observed particles, and how dynamic programming, \astar search, and reinforcement learning can be used to find the maximum likelihood clusterinng in this enormous search space. This work builds bridges with work in hierarchical clustering, statistics, combinatorial optmization, and reinforcement learning. 
\end{abstract}

%% file: tex/01_intro.tex
\vspace{-.5cm}
\section{Introduction}
\vspace{-.1cm}

Jets are the most copiously produced objects at the LHC and the subject of intense experimental and theoretical study.  Improvements to our understanding and treatment of jets can have a significant impact on the physics program of the LHC; however, various computational bottlenecks appear in this quest. Below we will discuss a few areas that show such computational bottlenecks and identify emerging computational techniques that may be able to address them. We hope that this may challenge some assumptions about the computational demands of simulation, reconstruction, and analysis of LHC data when jets are involved.

\vspace{-0.3cm}
\subsection{Reframing jet physics in probabilistic terms}\label{probFraming}
\vspace{-0.1cm}

Monte Carlo event generators (e.g. simulators like PYTHIA ~\cite{Sjostrand:2014zea}, Herwig ~\cite{Bellm:2015jjp}, and Sherpa ~\cite{Gleisberg:2008ta}) encode a physics model for the fragmentation and hadronization of quarks and gluons produced at colliders. In statistical and machine learning language, they are generative models for jets. Following the notation of Ref.~\cite{Brehmer:2018eca,Cranmer201912789}, we denote the parameters of the (Monte Carlo) simulation $\theta$, the observable output of the simulator $x$, and latent variables (aka Monte Carlo truth record or showering history) $z$. The simulators typically evolve the latent state  sequentially as a Markov process and model the physics of each splitting, clustering, etc. 
In the original parton showers, based on successive $1\to 2$ splittings, the joint likelihood for the parton shower can be expressed as:
\begin{equation}
p(x,z | \theta) = p(x | z_\textrm{leaves},\, \theta) \prod_{s \in \textrm{splittings}} p(z_{s,L}, z_{s,R} | z_{s,P}, \,\theta) \; ,
\label{eq:jet_markov}
\end{equation}
where  $z_{s,P}$, $z_{s,L}$ ($z_{s,R}$), are respectively the data needed to encode the state of the parent and left (right) children for the $s^\textrm{th}$ splitting and $z_\textrm{leaves}$ are the terminal leaves of the showering process. The hadronization and detector simulation fit in this framing as well, but we do not discuss it explicitly in this work.

We find it elucidating to reframe the following concepts in jet physics in probabilistic terms: 
\begin{itemize}    
    \item Joint likelihood for latent shower and observed constituents $p(x,z | \theta)$
    \item Marginal likelihood for observed constituents $p(x|\theta) = \int dz\,\, p(x,z | \theta)$
    \item Maximum likelihood showering history $\hat{z} = \text{argmax}_z \,p(x |z, \theta)$
    \item Maximum likelihood parameters for the model $\hat{\theta} = \text{argmax}_\theta \,p(x | \theta) = \text{argmax}_\theta \int dz\,\, p(x,z | \theta)$
    \item Posterior distribution on showering histories $p(z | x , \theta)$
\end{itemize}
A few challenges present themselves in this framing of jet physics. 

First of all, the joint likelihood $p(x,z | \theta)$ and the likelihood of individual splittings $p(z_{s,L}, z_{s,R} | z_{s,P},\, \theta)$ is not exposed in a way that is convient to access. The joint likelihood corresponds to  what is coded in PYTHIA ~\cite{Sjostrand:2014zea}, Herwig ~\cite{Bellm:2015jjp}, and Sherpa ~\cite{Gleisberg:2008ta}, but often in terms of accept-reject sampling and procedural code that does not explicitly expose the probabilities themselves. This motivates \Ginkgo, which provides convenient access to these quantities in a simplified parton shower.

Secondly, the joint likelihood $p(x,z | \theta)$ is not immediately of interest to experimentalists since the (latent) showering history $z$ is not observed. Quantities such as the marginal likelihood $p(x|\theta)$ and the maximum likelihood parameter $\hat{\theta}$ involve integration (sums) over all possible showering histories. The number of possible showering histories grows factorially with the number of jet constituents. This super-exponential growth in the number of showering histories is at the heart of many computational bottlenecks in jet physics, making the marginalization and maximization over the latent space $z$ of showering histories typically intractable.   

This paper is organized as follows. Below, we will review some common tasks in jet physics framed in these probabilistic terms. We will identify the computational challenges and the potential for emerging computational techniques to address them. In Section \ref{sec:Ginkgo} we will describe \Ginkgo's simplified probabilistic model for the parton shower. Finally, in Section \ref{ResearchWithGinkgo} we will review some of the recent research into these problems enabled with \Ginkgo.

\vspace{-0.3cm}
\subsubsection{Jet clustering}\label{sec:clustering}
\vspace{-0.1cm}

Jet reconstruction can be thought of as estimating the latent state (showering history) $z$ from the observed particles $x$. 
Traditionally, given a set of final state particles, jets are reconstructed using one of the generalized $k_t$ clustering algorithms~\cite{ Cacciari:2008gp,Catani:1993hr,Dokshitzer:1997in, Ellis:1993tq}. These algorithms sequentially cluster jet constituents by merging the closest pair based on the distance measure $d_{ij}^\alpha$,
\beq\label{eq:dij}
d_{ij}^\alpha =\text{min}(p_{ti}^{2 \alpha},p_{tj}^{2 \alpha}) \frac{\Delta R_{ij}^2}{R^2}
\eeq
where $\Delta R_{ij}$ is the angular distance for the pair $\langle i,j \rangle$, $R$ is a fixed value corresponding to the jet radius and  $\alpha = \{-1,0,1\}$ specifies the anti-$k_t$, Cambridge/Aachen (C/A) and $k_t$ algorithms, respectively. This traditional approach does not make explicit reference to the probability model for the jet, but intuitively the $k_t$ and C/A algorithms cluster together two constituents that are likely to have emerged from the same parent.\footnote{In contrast, the anti-$k_t$ algorithm focuses more on having desirable global properties for the jets than reconstructing a physically motivated showering history.} We can formulate the intuition for the $k_t$ and C/A algorithms as saying that the distance measure $d_{i,j}^\alpha$ is monotonically decreasing with the splitting likelihood $p(z_{s,L}, z_{s,R} | z_{s,P}, \,\theta)$.

In the probabilistic language, the natural goal is to find the most likely clustering $\hat{z}$. In that light, the generalized-$k_t$ algorithms are greedy algorithms for finding the maximum likelihood clustering. Greedy algorithms are not guaranteed to find the maximum likelihood clustering because they do not consider the tree globally, i.e. they find the best solution locally step by step. More sophisticated algorithms like beam search, which is used widely in natural language processing, look more than one step ahead and consider multiple possible clusterings in memory as they proceed. They are guaranteed to recover jet clusterings that are at least as good as the greedy algorithm, and can be expected to improve upon them. But to do this, one needs a way to score the combination of multiple clusterings. It is not clear how one would combine the distance measure $d_{i,j}^\alpha$ for two splittings, but there is a natural rule for combining the splitting likelihood $p(z_{s,L}, z_{s,R} | z_{s,P}, \,\theta)$ (i.e. their product). This is one of the advantages of the probabilistic formulation: it allows us to recast the objective of existing greedy clustering algorithms like $k_t$ as well as extend them to more sophisticated algorithms. 

In more general terms, jet clustering can be framed as a hierarchical clustering task. In that framing, the generalized-$k_t$ algorithms are considered (bottom-up) hierarchical agglomerative clustering (HAC) algorithms. 
In Section \ref{ResearchWithGinkgo} we will review the research in jet clustering enabled by \Ginkgo, including a novel trellis data structure and dynamic programming model, a novel $A^*$ search algorithm that makes use of a modified trellis data structure and a heuristic, as well as a few reinforcement learning algorithms including Monte Carlo Tree Search and Behavioral Cloning.

\begin{figure*}
\centering
\begin{minipage}[c]{0.25\textwidth}
 	 {
  \includegraphics[width=\linewidth]{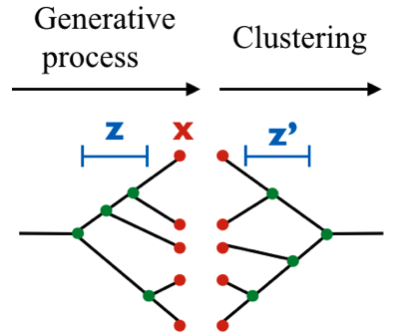}
	}
\end{minipage}
\hspace{1cm}
\begin{minipage}[c]{0.49\textwidth}
\caption{\small{Schematic representation of the tree structure of a jet generated with Ginkgo and the resulting tree for some clustering algorithm. For a given algorithm, $z$ labels the latent structure of the tree. The tree leaves $x$ are labeled in red and the inner nodes in green. 
}}
\end{minipage}
\label{fig:treestructure2}
\vspace{-0.6cm}
\end{figure*}

\vspace{-0.3cm}
\subsubsection{Tuning the parameters of the shower model}
\vspace{-0.1cm}
Monte Carlo tuning, can be thought of as estimating $\theta$ given a dataset of $\{x_i\}$. 
Ideally, if we wanted to fit (tune) the parameters $\theta$ of PYTHIA ~\cite{Sjostrand:2014zea}, Herwig ~\cite{Bellm:2015jjp}, and Sherpa ~\cite{Gleisberg:2008ta} (and we had infinite computing power), then we would compute the maximum likelihood $\hat{\theta}$ based on the high dimensional jet data $x$. Since we do not, we resort to tools like \texttt{Professor}~\cite{Buckley:2009bj}, which compare projections of complicated events to individual variables (marginal distributions), which is blind to various forms of mismodelling in the high-dimensional structure of the jets.  The marginalizaton over the latent space is implicit when forming the histograms of these individual differential distributions. The fact that tuning the generators is itself a bottleneck suppresses the motivation to add even more flexibility and parameters to the shower models, even if they might lead to more accurate description of the jets.

The first emerging technique in this direction is {\it likelihood-free inference} or {\it simulation-based inference} \cite{Brehmer:2018eca, Brehmer:2018kdj,  Brehmer:2018hga}. Recent progress in this direction includes likelihood-free inference methods \cite{ Brehmer:2018eca,Cranmer201912789,Brehmer:2018kdj, Brehmer:2018hga}. These methods approximate the intractable $p(x|\theta)$  using machine learning and bypass an explicit  marginalization over the latent state $z$. The techniques can exploit the joint likelihood $p(x ,z | \theta)$ if it is available. An implementation of these techniques for events simulated with Pythia was introduced in Refs.~\cite{Andreassen:2019nnm}. A closely related approach was outlined in Ref.~\cite{Louppe:2017pay}

An alternate approach to this problem would be to use probabilistic programming techniques to efficiently approximate the intractable integrals. The first prototypes of integrating probabilistic programming with the Monte Carlo generators (specifically Sherpa) was performed in Refs.~\cite{Baydin:2019fap, Baydin:2018npr}

\vspace{-0.3cm}
\subsubsection{Event Generation for events with large jet multiplicity}
\vspace{-0.1cm}
The enormous number of possible showering histories is a  bottleneck in the simulation of multijets events~\cite{Catani:2001cc, Lonnblad:2001iq} and shower deconstruction~\cite{Soper:2011cr, Soper:2012pb, FerreiradeLima:2016gcz}. When implementing the CKKW-L matching algorithm \cite{Catani:2001cc, Lonnblad:2001iq}, parton final states need to be reweighted with the corresponding Sudakov form factors of each history, $p(x,z | \theta)$.
The standard algorithm typically becomes infeasible for parton level configurations that exceed the complexity of $W/Z$+ 6 jet final state \cite{Hoeche:2019rti} due to the super-exponential growth in the number of clustering histories.

To ameliorate these bottlenecks we introduced in \cite{Greenberg:2020heb}, a novel data structure and algorithms, called hierarchical cluster trellis, that can be used to efficiently represent the distribution over trees. The trellis can be used to compute the marginal likelihood $p(x|\theta)$ or the exact maximum likelihood showering history $\hat{z}$ in time and memory proportional to the significantly smaller powerset of the number of jet constituents, i.e. $2^N$. In particular, we showed that the trellis allows us to perform these operations for larger values of $N$ where the naive iteration over the $(2N-3)!!$ trees is impractical.
Thus far the implementation is based on binary trees, $1 \rightarrow 2$ splittings; however, it is possible to extend the cluster trellis to consider $2 \rightarrow 3$ splittings required in the CKKW-L algorithm.

The trellis data structure also provides an efficient dynamic programming algorithm to find the maximum likelihood shower history $\hat{z}$, which provides a principled alternative to the generalized $k_t$ algorithms, that are based on a greedy sequential clustering algorithm as described in Sec.~\ref{sec:clustering}.

\vspace{-0.3cm}
\subsubsection{Simulating jet backgrounds in signal-rich regions of phase space}
\vspace{-0.1cm}
Simulating sufficient numbers of multijet background events is a  computational challenge due to the enormous rate of multijet events and their steeply falling spectra. The experimental collaborations have traditionally sliced the phase space into exclusive regions  (eg. based on the $p_T$ of the leading jet at parton level). This is an effective strategy for populating the tails of that distribution, but it is not effective for populating complicated phase space regions (eg. QCD events that satisfy the cut on a boosted top-tagger). Generating enough background Monte Carlo events in these signal-like regions of phase space is one of the computational challenges for the LHC and HL-LHC.

Denote passing an event selection cut with the indicator function $\mathbf{1}(x)$. In our probabilistic language, we are interested in efficiently sampling the showering histories $z \sim p(z | \mathbf{1}(x), \theta)$ so that we do not waste computing resources on the expensive detector simulation for events that won't satisfy the cuts. When the phase space regions are not aligned with parton level quantities, then we must perform importance sampling in the parton shower itself, and the ideal importance distribution would be the unfolded $p(z | \mathbf{1}(x), \theta)$, which is difficult to estimate when working with cuts based on complicated jet observables. 

Recent developments in probabilistic programming systems offer a potential way to address these challenges. Probabilistic programming systems provide tools for inferring the latent state of a simulator based on some observations (e.g., $p(z|x,\theta)$), and they use the simulator directly during inference. As mentioned above, the \texttt{pyprob} probabilistic programming system was integrated with the Sherpa event generator via the \texttt{ppx} protocol~\cite{Baydin:2019fap, Baydin:2018npr}. By instrumenting Sherpa with \texttt{ppx}, the \texttt{pyprob} system is able to bias the control flow of the event generator to perform advanced forms of importance sampling. In Refs.~\cite{Baydin:2019fap, Baydin:2018npr} a large recurrent neural network learned an efficient importance sampling distribution $q(z|x)$; however, the target was $\tau$ lepton decay instead of jet physics. More recently, we have instrumented our \Ginkgo~generator with \texttt{pyprob}, which we will describe below.

%% file: tex/02_model_description.tex
\vspace{0.6cm}
\section{Ginkgo: A simplified generative model for jets}\label{sec:Ginkgo}

At present, it is very hard to access the joint likelihood in state-of-the-art parton shower generators in full physics simulations, e.g. PYTHIA ~\cite{Sjostrand:2014zea}, Herwig ~\cite{Bellm:2015jjp}, and Sherpa ~\cite{Gleisberg:2008ta}. Also, typical implementations of parton showers involve 
sampling procedures that destroy the analytic control of the joint likelihood.
Thus, to aid in machine learning research for jet physics, a python package for a toy generative model of a parton shower, called \Ginkgo, was introduced in \cite{ToyJetsShowerPackage}.
Ginkgo has a tractable joint likelihood, and is as simple and easy to describe as possible but at the same time captures essential ingredients of parton shower generators in full physics simulations. It also ensures permutation invariance and momentum conservation. 
Ginkgo was designed to enable implementations of {\it probabilistic programming, differentiable programming, dynamic programming} and {\it variational inference}.
Within the analogy between jets and NLP, Ginkgo can be thought of as ground-truth parse trees with a known language model.

Ginkgo implements a recursive algorithm to generate a binary tree, where each node is represented by an energy-momentum vector and the leaves are the jet constituents.  
We want our model to represent the following features:
\begin{itemize}
\setlength{\itemsep}{-4pt}%
\item Momentum conservation: the total momentum of the jet (root of the tree) is obtained from adding the momentum of all of its constituents.
\item Running of the splitting scale: each splitting is characterized by a scale $t$ that decreases when evolving down the tree from root to leaves ($t$ is the invariant squared mass, $t = m^2$).
\end{itemize}

We also want our model to lead to a natural analogue of the generalized $k_t$ clustering algorithms~\cite{ Cacciari:2008gp,Catani:1993hr,Dokshitzer:1997in, Ellis:1993tq} for the generated jets. These algorithms are characterized by
\begin{itemize}
\setlength{\itemsep}{-4pt}%
\item Permutation invariance: the jet momentum should be invariant with respect to the order in which we cluster its constituents.
\item Distance measure: the angular separation between two jet constituents is typically used as a distance measure among them. In particular, traditional jet clustering algorithms are based on a measure given by 
$d_{ij} \propto  \Delta R_{ij}^2$
where $\Delta R_{ij}$ is the angular separation between two particles.
\end{itemize}

\vspace{-0.3cm}
\subsection{The generative process}\label{Generatiion}
\vspace{-0.1cm}
The generative process depends on the following input parameters:
\begin{itemize}
\setlength{\itemsep}{-3pt}%
\item $p_0^\mu$: four-momentum of the jet (input value for the root node of the tree).
\item $t_0$: initial squared mass. 
\item $t_\text{cut}$: cut-off squared mass to stop the showering process. 
\item $\lambda$: decaying rate for the exponential distribution.
\end{itemize}

During the generative process, starting from the root of the tree, each parent node is split, generating a left (L) and a right (R) child. At each splitting we sample squared invariant masses for the children, $t_L, t_R$ from a decaying exponential. We require the constraint $\sqrt{t_L} + \sqrt{t_R} < \sqrt{t_P}$, where $\sqrt{t_P}$ is the parent mass. Then we implement a 2-body decay in the parent center-of-mass frame. The children direction is obtained by uniformly sampling a unit vector on the 2-sphere (in the parent center-of-mass frame the children move in opposite directions). Finally, we apply a Lorentz boost to the lab frame, to obtain the 4 dimensional vector $p_{\mu}=(E, p_x, p_y, p_z) $ that characterizes each node.
This prescription ensures {\it momentum conservation} and {\it permutation invariance}.
Next, we outline the splitting of a node as follows.
\begin{enumerate}
\setlength{\itemsep}{-3pt}%
\item Draw $ t_{\text{L}}$ and  $t_{\text{R}}$ from an exponential distribution,
\vspace{-0.2cm}
\bea \label{eq:exponential1}
t_L \sim f(t | \lambda, t_{\text{P}})=\frac{1}{1-e^{- \lambda}} \frac{\lambda}{t_{\text{P}}} e^{- \frac{\lambda}{ t_{\text{P}}} t} 
\eea

\vspace{-0.8cm}
\bea \label{eq:exponential2}
t_R \sim f(t | \lambda, t_{\text{P}}, t_L)= \frac{1}{1-e^{- \lambda}} \frac{\lambda}{ (\sqrt{t_{\text{P}}}-\sqrt{t_L})^2} e^{- \frac{\lambda}{ (\sqrt{t_{\text{P}}}-\sqrt{t_L})^2} t} 
\eea
\vspace{-0.5cm}

and  define $m_L=\sqrt{t_\text{L}}$ and $m_R=\sqrt{t_\text{R}}$. We apply a veto on sampled values where $t_L  \geqslant t_{\text{P}}$ and $t_R \geqslant (\sqrt{t_{\text{P}}}-\sqrt{t_L})^2$. For inference, given two particles, we assign $t_L \rightarrow \text{max}\{t_L, t_R\} $ and $t_R \rightarrow \text{min}\{t_L, t_R\} $.
\item Compute a 2-body decay in the parent rest frame.
\item Apply a Lorentz boost to each of the children, with $\gamma=\frac{E_p}{\sqrt{t_P}}$ and $\gamma \beta = |\vec{p}_p|/\sqrt{t_P}$. 
\item If $t_{\text{L}}$ ($t_{R}$) is greater than  $t_\text{cut}$ repeat the process.
\end{enumerate}
The algorithm is outlined in Algorithm 1. After running, the final binary tree for the jet is obtained.
\begin{algorithm}
    \SetKwInOut{Input}{Input}
    \SetKwInOut{Output}{Output}

    function NodeProcessing $(p^\mu_{\text{p}}, t_\text{P}, t_\text{cut}, \lambda, \text{tree})$\\
    \Input{parent momentum $\vec{p}_{\text{p}}$, parent mass squared $t_\text{p}$, cut-off mass squared $t_\text{cut}$, rate for the exponential distribution $\lambda$, binary tree $tree$}
    \Indp
    Add parent node to tree. \\
    \If{$t_\text{p} > t_\text{cut}$}
      {
      	Sample $t_L$ and $t_R$ from the decaying exponential distribution.\\
	Sample a unit vector from a uniform distribution over the 2-sphere.\\
	Compute the 2-body decay of the parent node in the parent rest frame.\\
	Apply a Lorentz boost to the lab frame to each child.\\
	NodeProcessing $(p^\mu_{\text{p}}, t_\text{L}, t_\text{cut}, \lambda, \text{tree})$\\
	NodeProcessing $(p^\mu_{\text{p}}, t_\text{R}, t_\text{cut}, \lambda, \text{tree})$\\
      }
    \caption{Toy Parton Shower Generator}
\end{algorithm}
\vspace{-0.3cm}
\subsection{Reconstruction: The Likelihood for a Proposed Jet Clustering}
\vspace{-0.1cm}
In addition to the generative model described above, which is used for generating data with Monte Carlo, we also need to be able to assign a likelihood value to a proposed jet clustering. To do this we use the same general form for the jet's likelihood based on a product of likelihoods over each splitting as in Eq.~\ref{eq:jet_markov}. In order to evaluate this we need to first reconstruct the parent from the left and right children. Then we use the same equations described above (Eq.~\ref{eq:exponential1} and \ref{eq:exponential2}) for the splitting probabilities that are used in the generative model. The~\Ginkgo library provides functions to evaluate the joint likelihood $p(x,z | \lambda, t_\textrm{cut})$ of any proposed hierarchical clustering of the observed final state particles.

%% file: tex/03_greedy_beamsearch.tex
\vspace{-0.3cm}
\subsection{Greedy and beam search algorithms}\label{sec:Greedy_BS}
\vspace{-0.1cm}

As described in Sec.~\ref{sec:clustering}, we can reframe the goal of jet clustering as finding the maximum likelihood estimate (MLE) for the latent structure of a jet, given a set of constituents (leaves). Different algorithms will return different tree-like hierarchical clusterings $z_{\text{shower}}$, and we can compare the performance of various algorithms. We study approximate solutions for bottom-up agglomerative clustering like the generalized-$k_t$ algorithms (which are a class of greedy algorithms that locally maximize the likelihood at each step in the clustering process) and beam search (which maximizes the likelihood of multiple steps before choosing what to cluster).

\begin{figure*}
\centering
\begin{minipage}[c]{0.49\textwidth}
 	 {
	  \includegraphics[width=\textwidth]{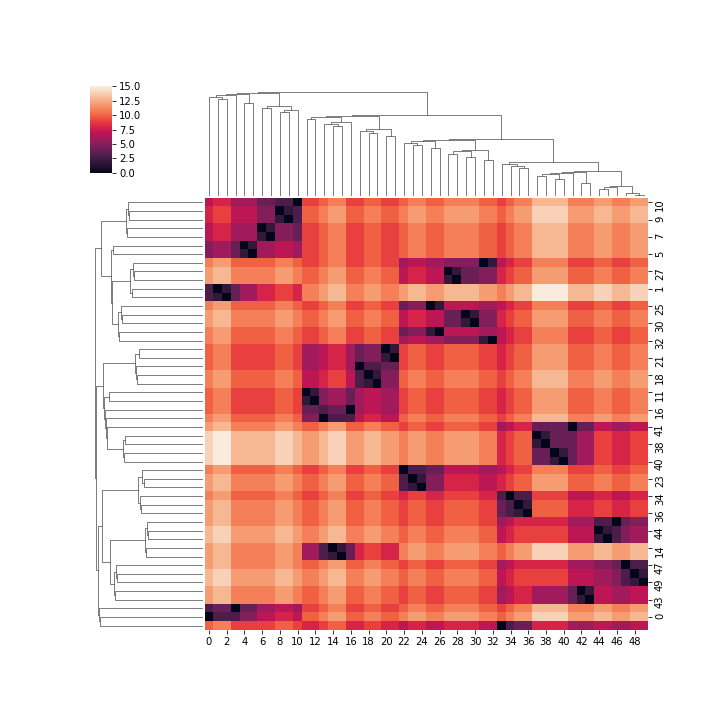}
	}
\end{minipage}
\begin{minipage}[c]{0.49\textwidth}
 	 {
	  \includegraphics[width=\textwidth]{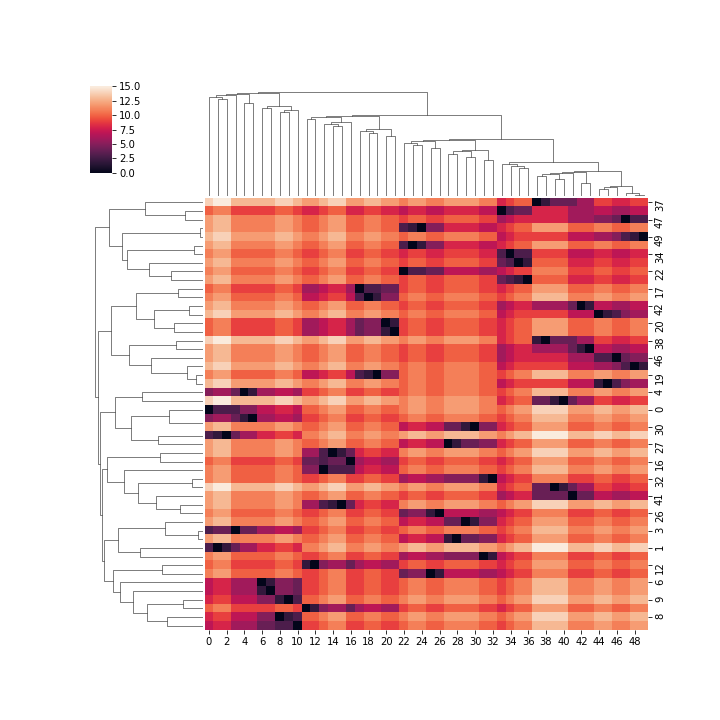}
	}
\end{minipage}
\vspace{-1.2cm}
\caption{\small{2D heat clustermap visualizations of~\Ginkgo~jets,  where the leaves ordering corresponds to the order to access them when traversing the truth tree (columns) and each clustering algorithm (rows), where we show beam search (left) and $k_t$ (right).
}}
\label{fig:2Dclustermap70}
\vspace{-0.5cm}
\end{figure*}

We provide implementations of these algorithms on~\Ginkgo~jets in \cite{TreeAlgorithms}.
We also developed a visualization package in ~\cite{VisualizeBinaryTrees} and show examples below. 
In Fig. \ref{fig:2Dclustermap70} we show 2D heat clustermaps where the color scale specifies the total number of steps needed to connect any two leaves through their closest common ancestor using the truth-level jet tree.
The better the truth tree latent structure is reconstructed, the more the heat map structure looks block diagonal.

\vspace{-0.3cm}

%% file: tex/04_hierarchical-clustering.tex
\section{Examples of Research Enabled with Ginkgo }\label{ResearchWithGinkgo}

In this section we highlight some of recent research that has been enabled with \Ginkgo. 

\vspace{-1em}

\subsection{Hierarchical Cluster Trellis Algorithm}\label{Uncertainty in Clustering}
\vspace{-0.1cm}

Jet clustering in high-energy physics is a  siloed sub-field of research, which is ironic given that hierarchical clustering is a common task in many areas of science and can be effectively abstracted. Hierarchical clustering is 
often used to discover meaningful structures, such as phylogenetic trees of organisms \cite{kraskov2005hierarchical}, taxonomies of concepts \cite{cimiano2005learning}, subtypes of cancer \cite{sorlie2001gene}.

We define a hierarchical clustering as a recursive splitting of a dataset of elements, $\dataset = \{x_i\}_{i=1}^N$ into subsets until reaching singletons, e.g. leaves of a binary tree. 
This can equivalently be viewed as starting with the set of singletons and repeatedly taking the union of sets until reaching the entire dataset.

The authors of Ref.~\cite{Greenberg:2020heb} consider an energy-based probabilistic model for hierarchical clustering. 
The model is based
on measuring the compatibility of each pair of sibling nodes, described by a potential function $\EfunS: 2^{\dataset} \times 2^{\dataset} \rightarrow \RR^{+}$. We also denote the potential function for a hierarchical clustering $\tree$ and dataset $\dataset$ as 
$\EfunT(\dataset | \tree)$. Then, the probability of $\tree$ for the dataset
   $\dataset$, $P(\tree|\dataset)$, is equal to the unnormalized potential
  of $\tree$ normalized by the partition function (marginal likelihood),
  $Z(\dataset)$:
\vspace{-0.2cm}
  \begin{equation}
      P(\tree|\dataset) = \frac{\EfunT(\dataset | \tree)}{Z({\dataset})} \,\,\,\,\,\,  \text{ with } \,\,\,\,\,\, \EfunT(\dataset | \tree) = \prod_{X_L,X_R \in \textsf{siblings}(\tree)} \EfunS(X_L,X_R)
  \end{equation}
where the partition function is given by $    Z({\dataset}) = \sum_{\tree \in \treeset({\dataset})} {\EfunT(\dataset | \tree)}. \label{eq:Z}$

Next, they define MAP hierarchy as the maximum likelihood hierarchical clustering given a dataset $\dataset$.
Exactly performing inference on the MAP hierarchy and finding the partition
function by enumerating all hierarchical clusterings over $N$ elements is exceptionally difficult because the number of hierarchies grows extremely rapidly, namely $\exactNumClusterings$~\cite{callan2009combinatorial,DaleMoonCatalanSets}.

To overcome the computational burden, a cluster trellis data structure for hierarchical clustering was introduced in \cite{Greenberg:2020heb}. The trellis computes these quantities in the $\mathcal{O}(3^{N})$ time, without having to iterate over each possible hierarchy. While still exponential,  this is  feasible in regimes where enumerating all possible trees would be infeasible, and is to our knowledge the fastest exact MAP/partition function result, making practical \emph{exact} inference for datasets on the order of 20 points ($\sim 3\times10^9$ operations vs $\sim 10^{22}$ trees) or fewer.

%% file: tex/05_trellis.tex
We briefly review novel dynamic-programming algorithms for \emph{exact} (and approx.) inference in hierarchical clustering introduced in~\cite{Greenberg:2020heb}. The trellis allows us to compute the {\bf partition function} $Z(\dataset)$ and  {\bf MAP inference}, i.e. find the maximum likelihood tree structure. The Cluster Trellis package is available at \href{https://github.com/SebastianMacaluso/ClusterTrellis}{github.com/SebastianMacaluso/ClusterTrellis}. Each node in the trellis corresponds to all subsets of elements (jet constituents). A schematic representation and the assignment between nodes in a binary tree and nodes in the trellis is shown in Fig. 6.
\begin{figure*}[htb]
\centering
\begin{minipage}[c]{0.5\textwidth}
 	 {
\includegraphics[width=\columnwidth]{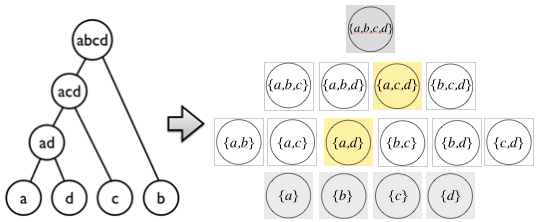}
	}
\end{minipage}
\hspace{0.4cm}
\begin{minipage}[c]{0.4\textwidth}
\caption{\small{Schematic representation of the trellis and node assignment between the trellis and a binary tree.
}}
\end{minipage}
\label{fig:sparseTrellis}
\end{figure*}

{\bf Computing the Partition Function.} 
The partition function, $Z({\dataset})$, for every node in the trellis is computed in order (in a bottom-up approach), memoizing the partial value at each node.

{\bf Computing the Maximum Likelihood Hierarchical Clustering.} 
The MAP hierarchy for dataset $\dataset$, $\treestar(\dataset)$, is $\treestar(\dataset) = \argmax_{\tree \in \treeset(\dataset)}{P(\tree|\dataset)} = \argmax_{\tree \in \treeset(\dataset)}{\EfunT(\tree)}$. This is also computed in order in a bottom-up approach.

{\bf Sampling from the Posterior Distribution.}\label{sec:Treesampling}
Drawing samples from the true posterior distribution 
$P(\tree|\dataset)$ is also difficult because of the 
extremely large number of trees. However, there is a sampling procedure implemented using the trellis which gives samples from the exact true posterior without enumerating all possible hierarchies.

\subsubsection{Sparse Cluster Trellis}\label{sec:sparsetrellis}
The authors of Ref.~\cite{Greenberg:2020heb} also introduced a sparse trellis data structure, which allows the algorithms to scale to larger datasets by controlling the sparsity index, i.e. the fraction of the total number of possible clusterings being considered.
Most clusterings have likelihood values orders of magnitude smaller than the MAP clustering making their contribution to the partition function negligible. 
As a result, if we build a sparse trellis that considers the most relevant hierarchies, we could find approximate solutions for inference
in datasets where implementing the full trellis is not feasible.
The sparse trellis can be constructed from samples (e.g., ground truth from a simulator, greedy, or beam search trees) or randomly sampling pairwise splittings for the children of a node.

\subsubsection{Results}
\begin{figure*}
\centering
\begin{minipage}[c]{0.42\textwidth}
        \centering
        \includegraphics[width=\columnwidth]{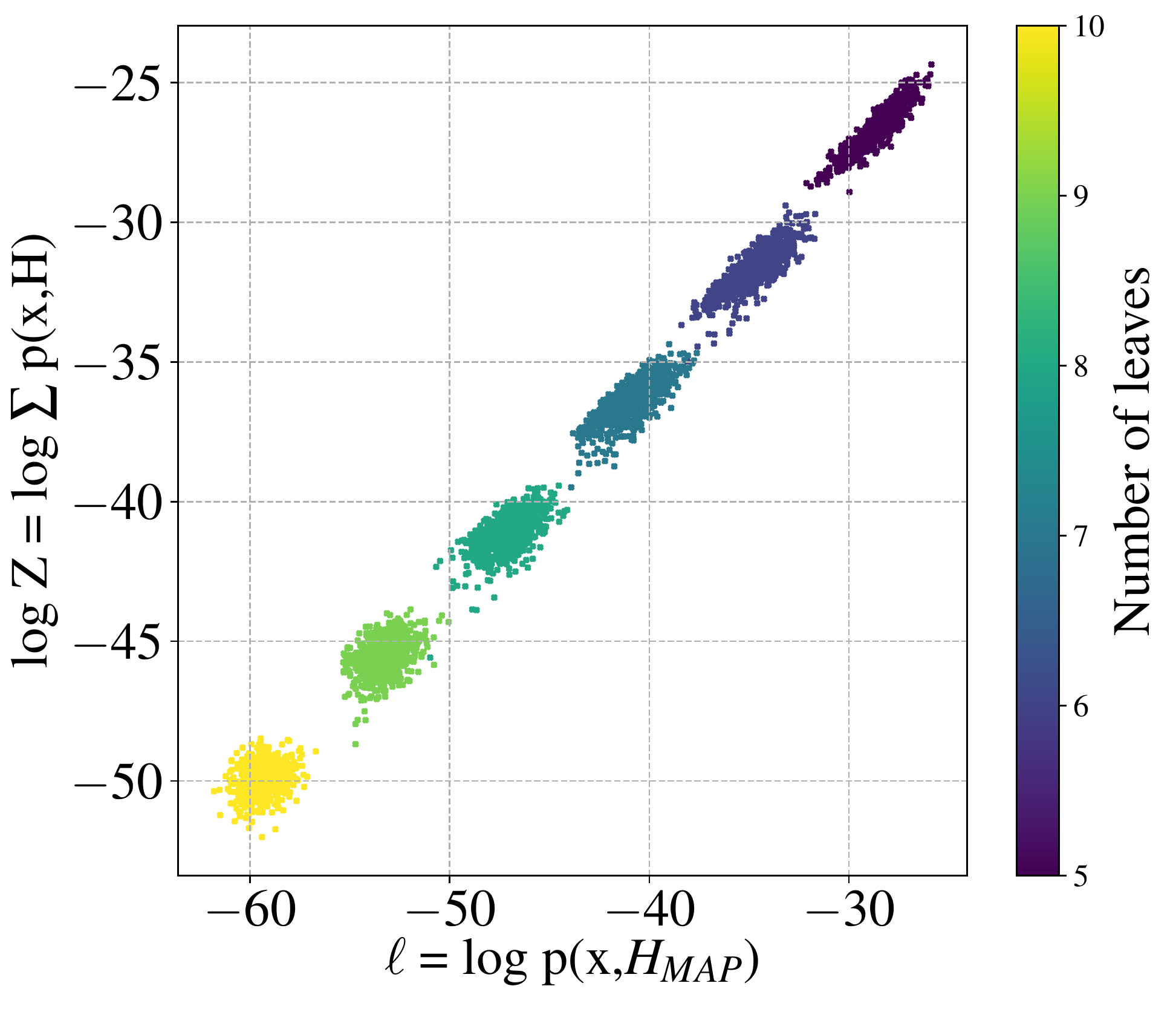}
\end{minipage}
\begin{minipage}[c]{0.42\textwidth}
        \centering
        \includegraphics[width=\columnwidth]{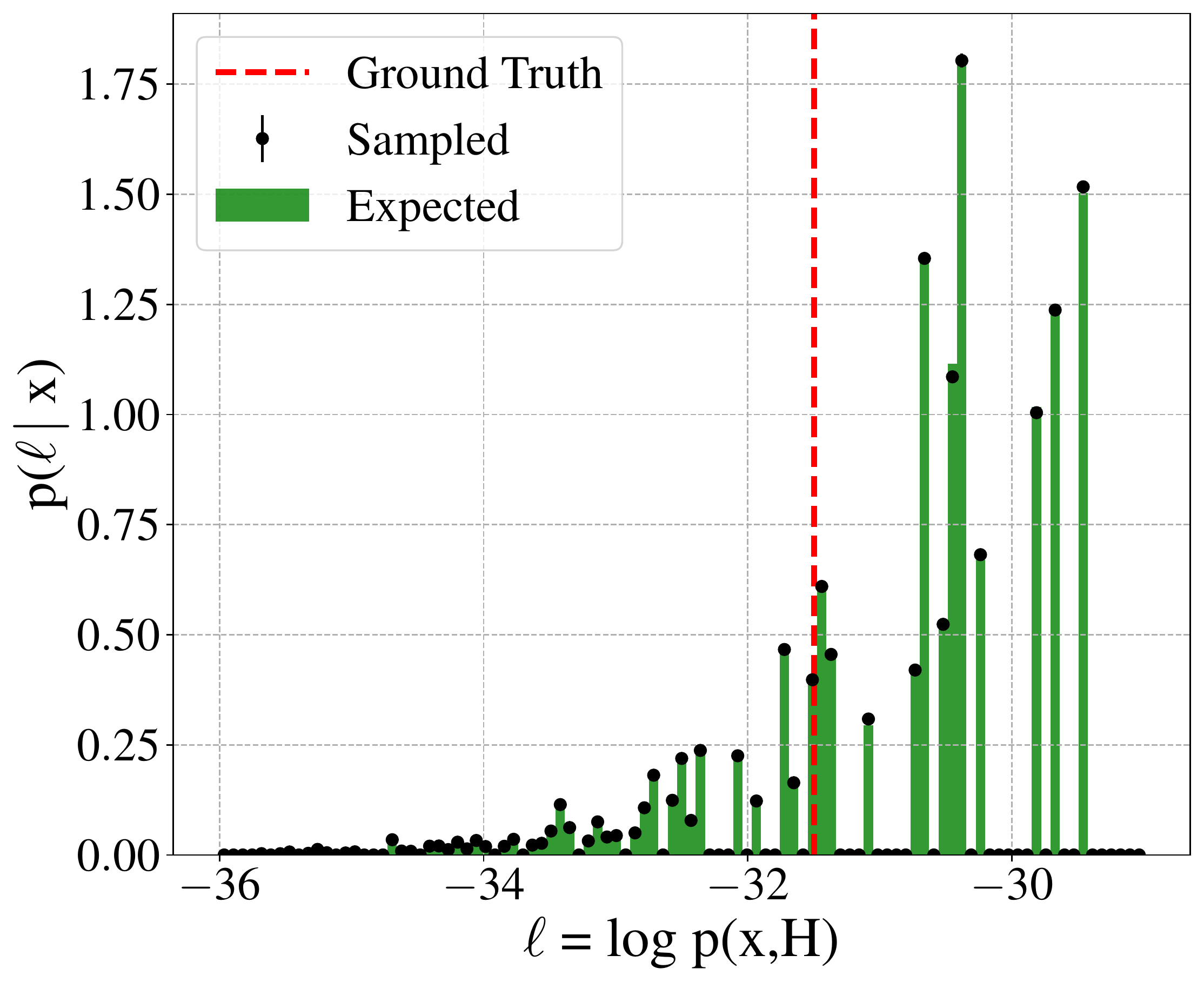}
\end{minipage}
\vspace{-0.3cm}
        \caption{\small{Left: scatter plot of the partition function $Z$  vs. the maximum log likelihood for a Ginkgo dataset, with up to 10 jet constituents. The color indicates the number of leaves of each hierarchical clustering. Right: comparison of the posterior distribution for  a specific jet with five leaves from sampling $10^5$ hierarchies (black dots with small error bars) and expected posterior distribution (in green). 
        The log likelihood for the ground truth tree is a vertical dashed red line.
        }}
        \label{fig:Z_Posterior}
\end{figure*}

In Fig. \ref{fig:Z_Posterior} (left) we show the  partition function versus the MAP hierarchy for each set of leaves from a Ginkgo dataset. 
Figure \ref{fig:Z_Posterior} (right) shows the results from sampling $10^5$ hierarchies (black dots) and the expected distribution.
Figure 5 shows the performance of the sparse trellis to calculate the MAP values on a set of 100 Ginkgo jets with 9 leaves. 
Even though beam search has a good performance for trees with a small number of leaves, we see that both sparse trellises quickly improve over beam search, with a sparsity index of only about 2\%. 
\begin{figure*}
\centering
\begin{minipage}[c]{0.35\textwidth}
 	 {
\includegraphics[width=\columnwidth]{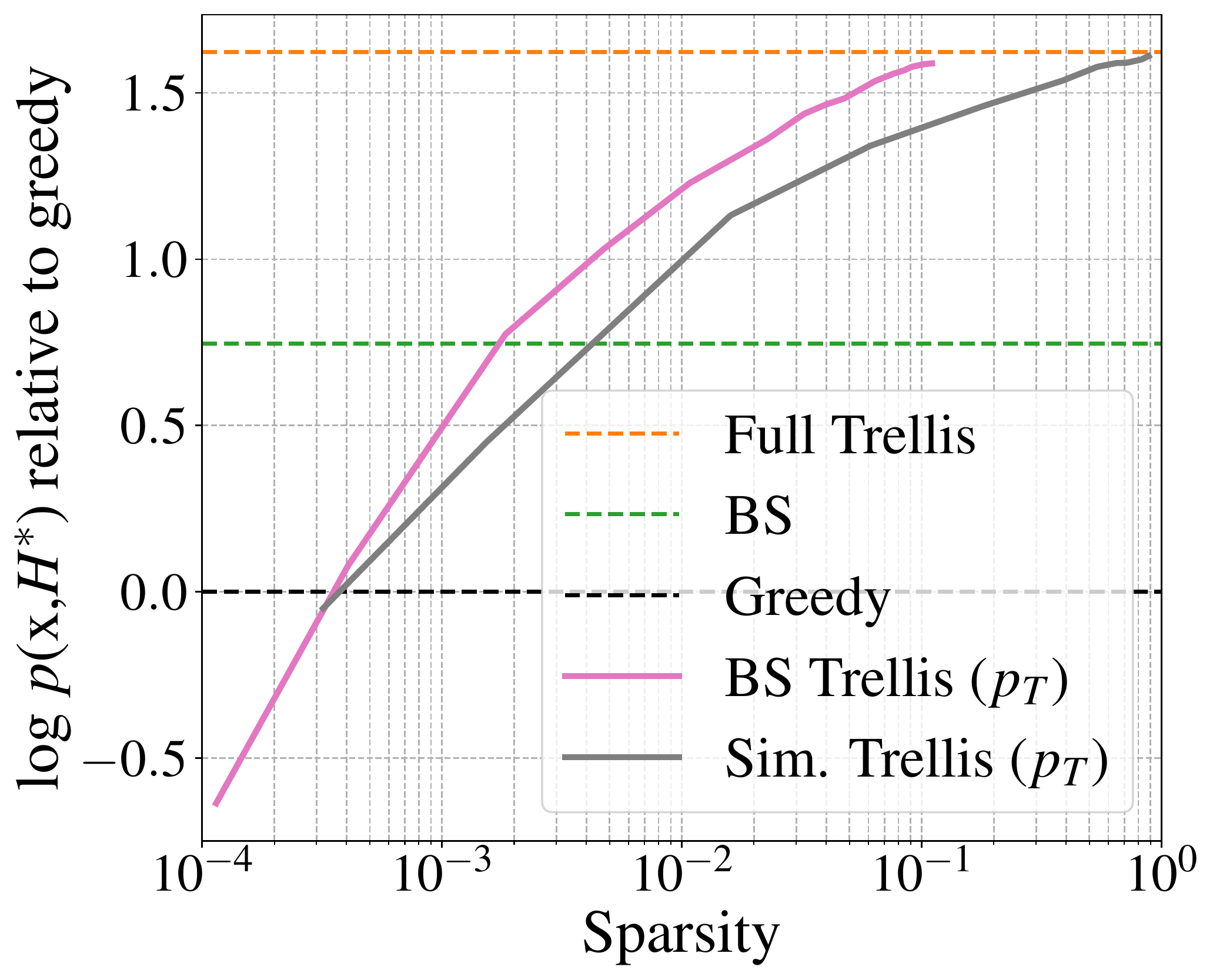}
	}
\end{minipage}
\hspace{1cm}
\begin{minipage}[c]{0.49\textwidth}
\caption{\small{Trellises MAP hierarchy log likelihood (values are relative to the greedy algorithm) vs their sparsity. Each value corresponds to the mean over 100 trees of a test dataset. We show the Simulator (Sim.) and the Beam Search (BS) trellises.  
}}
\end{minipage}
\label{fig:MAPvsSparsity}
\vspace{-0.8cm}
\end{figure*}

%% file: tex/06_ginkgo_RL.tex
\subsection{Hierarchical clustering through reinforcement learning}\label{sec:RL}

\vspace{-0.1cm}
In this section we review results from ~\cite{Brehmer:2020brs} that cast hierarchical clustering as a Markov Decision Process (MDP) and adapted reinforcement learning algorithms to solve it. 
In particular,  Monte-Carlo Tree Search (MCTS) guided by a neural network policy was adapted to the problem of jet clustering. This approach closely follows the AlphaZero algorithm ~\cite{silver2016mastering, silver2017mastering, silver2017mastering2}, which achieved superhuman performance in a range of board games, demonstrating its ability to efficiently search large combinatorial spaces. 
While (model-free) RL methods have been used in the context of jet grooming, i.e. pruning an existing tree to remove certain backgrounds~\cite{Carrazza:2019efs}, they have not yet been used for clustering, that is, the construction of the binary tree itself. 

\vspace{-0.3cm}
\subsubsection{Jet clustering as a Markov Decision Process}
\vspace{-0.1cm}

The authors of Ref.~\cite{Brehmer:2020brs} used the ingredients of \Ginkgo\ to recast the problem of clustering as an MDP, which is defined by the quartet $(\mathcal{S}, \mathcal{A}, P, R)$:
 \begin{itemize}
     \item The state space $\mathcal{S}$ is given by all possible particle sets at any given point during the clustering process, $s = z_t$.
     \item The actions $\mathcal{A}$ are the choice of two particles $a=(i, j)$ with $1 \leq i < j \leq n_t$ to be merged.
     \item The state transitions $P$ are deterministic and update $z_t$ to $z_{t - 1}$ by replacing the particles $p_{t,i}$ and $p_{t,j}$ with a parent $p_{t-1,i} = p_{t,i} + p_{t,j}$. All other particles are left unchanged, each state transition thus reduces the number of particles by one.
     \item The rewards $R$ are the splitting probabilities, $R(s=z_t, a=(i,j)) = \log p_s(z_{t} | z_{t-1}(i,j))$.
     \item The MDP is episodic and terminates when only a single particle is left.
 \end{itemize}

An agent solves the jet clustering problem by first considering the state of all observed, final-state particles and choosing which two to merge into a parent. It receives the log likelihood of this splitting as reward. Next, it considers the reduced set of particles where the two chosen particles have been replaced by their proposed parent, chooses the next pair of particles to merge, and so on. Rolling out an episode leads to a proposed clustering tree $z = \{z_1, \dots, z_N\}$, with the total received reward being equal to the log likelihood of this tree following Eq.~\eqref{eq:jet_markov}. 

The formulation of jet clustering as an MDP allows us to use any (model-free) reinforcement learning (RL) algorithm to tackle it. Since the state transition model is known (and deterministic), they instead use in ~\cite{Brehmer:2020brs}, a model-based planning approach to leverage this knowledge. They chose Monte Carlo Tree Search (MCTS)~\cite{silver2016mastering}, which builds a search tree over possible clusterings $z$ by rolling out a number of clusterings. 
In addition, the authors considered a clustering algorithm based on imitation learning, specifically Behavioral Cloning (BC): a policy $\pi$ is trained to imitate the actions that reconstruct the true trees, which we can extract from the generative model, by maximizing $\log \pi(s, a_\mathrm{truth})$.

\vspace{-0.3cm}

\subsubsection{Results}
\begin{figure}%
\vspace{-0.5cm}
    \centering%
    \includegraphics[width=0.35\linewidth]{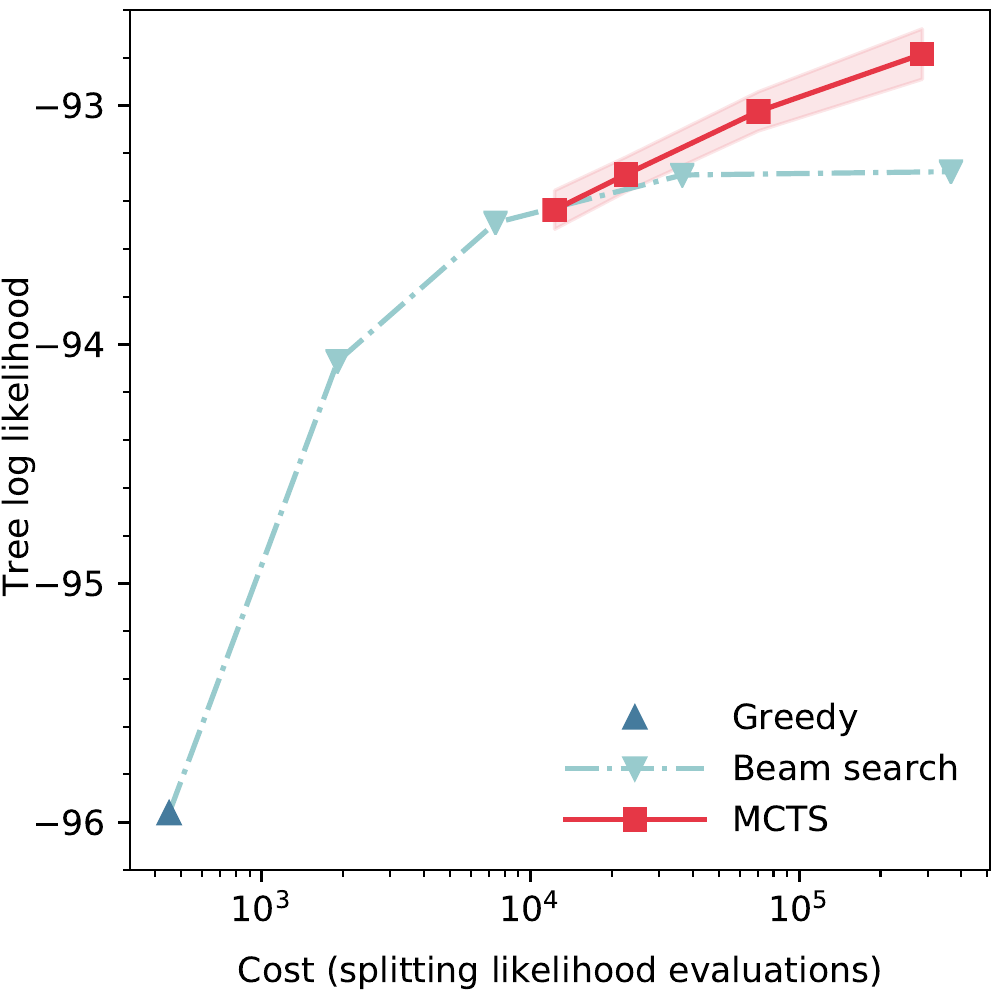}%
    \hspace*{0.04\linewidth}%
    \includegraphics[width=0.35\linewidth]{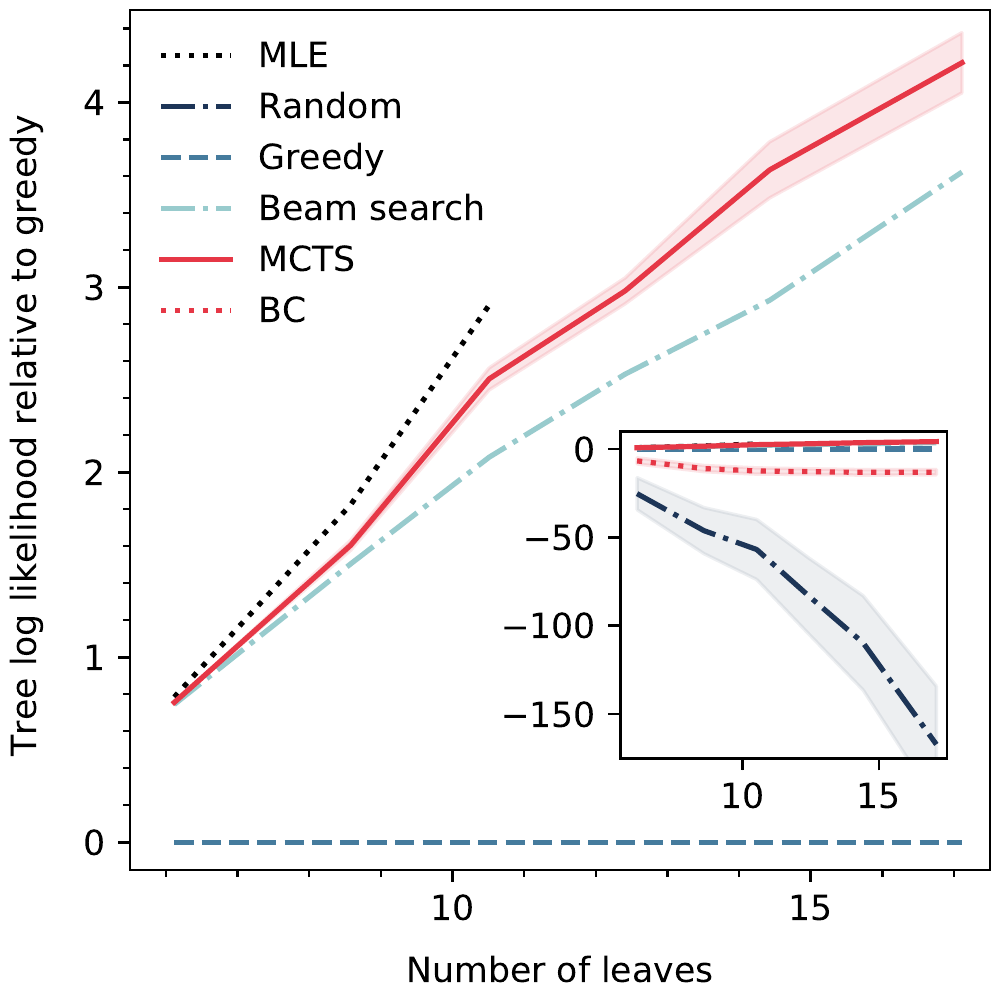}%
    \vspace{-0.3cm}
    \caption{Mean log likelihood of clustered jets (larger is better). \textbf{Left}: against the computational cost, measured as the number of evaluations of the splitting likelihood $p_s$ required by the different algorithms.  \textbf{Right}: as a function of the number of final-state particles (leaves of the tree), using the best-performing (and most computationally expensive) hyperparameter setup for each algorithm. MCTS (solid, red) gives the highest-quality jet clusterings.}%
    \label{fig:results}%
    \vspace{-0.7cm}
\end{figure}

We present a comparison of the different clustering algorithms on a dataset of Ginkgo jets, taken from ~\cite{Brehmer:2020brs}.
They compare MCTS and BC agents to a greedy algorithm, a beam search algorithm that maintains the $b=1000$ most likely clusterings, and a random policy. For jets with a small number of final-state particles we also compute the trellis exact maximum likelihood tree (MLE) following Ref.~\cite{Greenberg:2020heb}.
Fig.~\ref{fig:results} shows the log likelihood of the clustering against its computational cost (left) and against the number of final-state particles (right). While the greedy and beam search baselines lead to a robust performance at low computational cost, MCTS planning can generate hierarchical clusterings of a markedly higher likelihood. This advantage is more pronounced at larger number of final-state particles, showing that MCTS can explore large combinatorial spaces better than these baselines. 

%% file: tex/07_a_star.tex
\subsection{Hierarchical clustering using \astar}\label{sec:astar}

In this section we review an algorithm introduced in Ref. \cite{Greenberg:2021nrg}, that combines \astar search with a novel trellis data structure to overcome the prohibitively large search space of hierarchical clusterings, namely (2N-3)!! hierarchies for N elements. This method can be applied to the same type of energy-based probabilistic models for hierarchical clustering introduced in Ref.~\cite{Greenberg:2020heb}. 

\paragraph{\astar Search.}
\astar is a best-first search algorithm for finding the minimum cost path between a starting node and a goal node in a weighted graph. Following canonical notation, the function $f(n) = g(n) + h(n)$ determines the most promising node to search next, where $g(n)$ is the cost of the path up from the start to node $n$ and $h(n)$ is a heuristic estimating the cost of traveling from $n$ to a goal. Also, when the heuristic $h(\cdot)$ is \emph{admissible} (always under estimates cost), \astar is admissible and provides an optimal solution. 

In order to redefine clustering as a search problem, we need to define the search space, the graph being searched over, and the goal states. Each state in the search space is defined as a partial hierarchical clustering, which is a subset of some hierarchical clustering of the elements, e.g. a sub-tree.
The fundamental challenge of applying \astar to clustering
is the massive state space. Na\"ive representations of the \astar state space and frontier require explicitly storing every tree structure, potentially growing to be at least the number of binary trees $(2n-3)!!$.
To overcome this, in Ref.~\cite{Greenberg:2021nrg} an approach was proposed using a \emph{cluster trellis} to (1) compactly encode states in the space
of hierarchical clusterings (as paths from the root to the leaves of the trellis), and (2) compactly represent the search frontier (as nested priority queues). This method, reduces the super-exponential space and time required for \astar to find the exact MAP hierarchy to exponential in the worst case.

\paragraph{Aproximate \astar.}
An approx. version of \astar, feasible for a much greater number of elements (jet constituents), was also implemented in Ref.~\cite{Greenberg:2021nrg}. In this case, a sparse trellis (one with missing nodes and/or edges) is initialized using beam search. 
Then, (1) \astar is run over this sparse trellis, (2) the sparse trellis is extended by adding nodes, and edges corresponding to child / parent relationships while running \astar at exploration time for each node explored during \astar search, (3) \astar is run iteratively, obtaining a solution and then running \astar again, extending the trellis further between or during subsequent iterations.

\subsubsection{Results}

\begin{figure}
    \centering
    \includegraphics[width=0.65\textwidth]{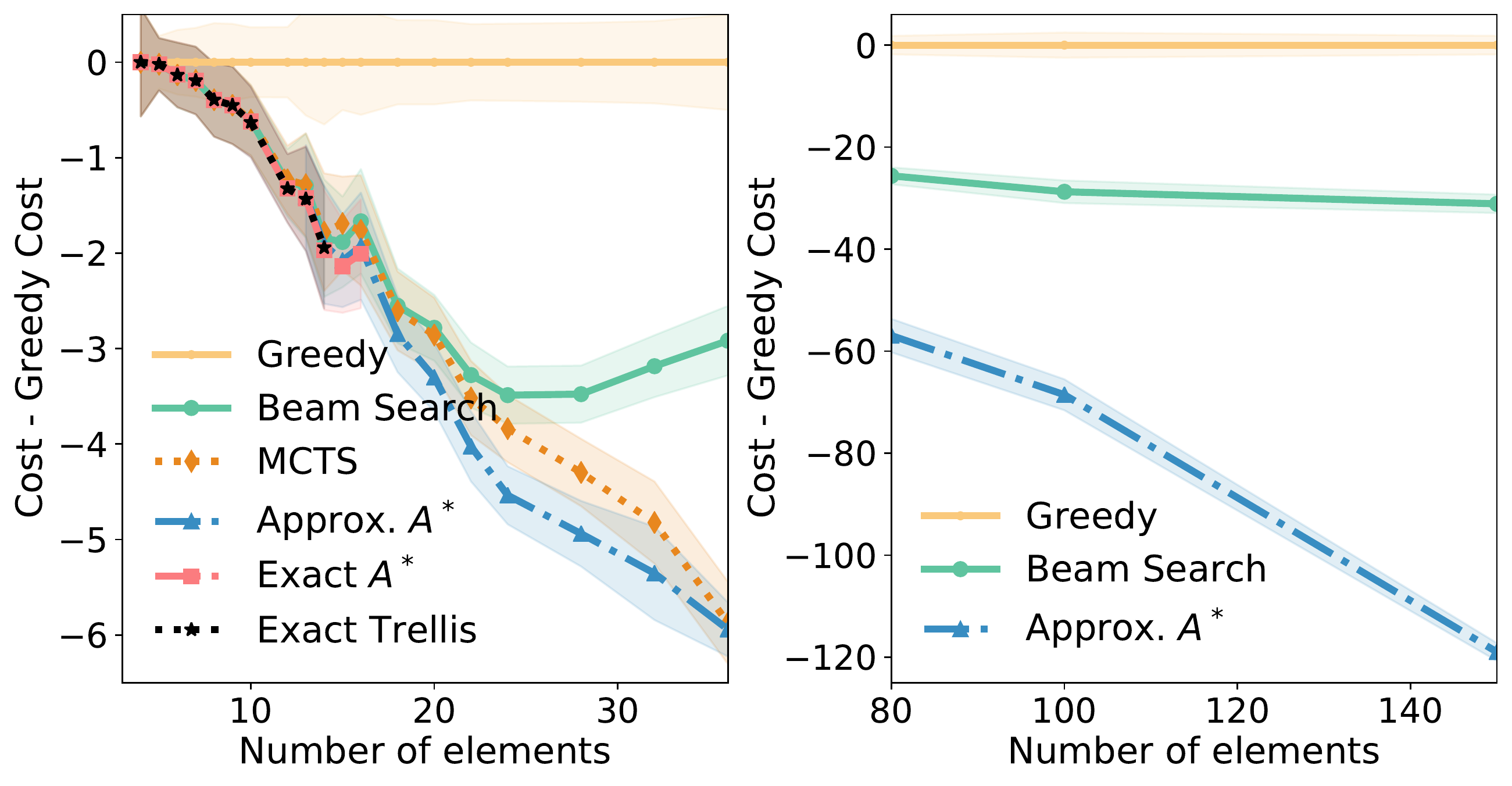}
    \vspace{-0.3cm}
    \caption{\small{\textbf{Jet Physics Results.} Cost (Neg. log. likelihood)  for the MAP hierarchy vs number of elements (jet constituents) of each algorithm on Ginkgo jets minus the cost for greedy (lower values are better solutions). We see that both exact and approx. \astar greatly improve over beam search and greedy. Though MCTS provides a strong baseline for small jets where it is feasible to implement it (left), \astar still shows an improvement over it.}}
    \label{fig:ginkgoCost}
\end{figure}

Figure \ref{fig:ginkgoCost} shows a comparison of the MAP values of the proposed exact and approximate algorithms using \astar with benchmark algorithms (greedy, beam search, MCTS, and exact trellis), on Ginkgo jets \footnote{These algorithms can be run from \href{https://github.com/SebastianMacaluso/HCmanager}{github.com/SebastianMacaluso/HCmanager}.}.
Approximate versions of \astar allow the algorithm to handle jets with many more constituents while significantly improving over beam search and greedy baselines. MCTS provides a strong baseline for small datasets, where it is feasible to implement it. However, \astar shows an improvement over MCTS while also being feasible for jets with more constituents.

%% file: tex/08_probabilistic_programming.tex
\vspace{-0.3cm}
\subsection{Probabilistic Programming}\label{sec:probprog}
\vspace{-0.1cm}

The Monte Carlo simulators implicitly describe the complicated distribution $p(x,z|\theta)$ and implement sampling through random number generators. Probabilistic programming extends this functionality with the ability to condition on the values of some of the random variables $x$ or $z$~\cite{gordon2014probabilistic}, and it achieves this by hijacking the random number generators. This controlling inference algorithm uses those hooks to bias the simulator towards the desired output (e.g. importance sampling~\cite{le2017inference}) or through Monte Carlo sampling. In particular, it provides the ability to sample latent variables conditioned on observations, i.e. $z\sim p(z|x,\theta)$, and observations conditioned on latent variables, i.e. $x\sim p(x|z, \theta)$. For example, this technique can be used to efficiently sample the tails of backgrounds in signal-rich regions of phase space $z \sim p(z | \mathbf{1}(x), \theta)$.

\begin{figure*}
\centering
\begin{minipage}[c]{0.35\textwidth}
 	 {
    \includegraphics[width=\linewidth]{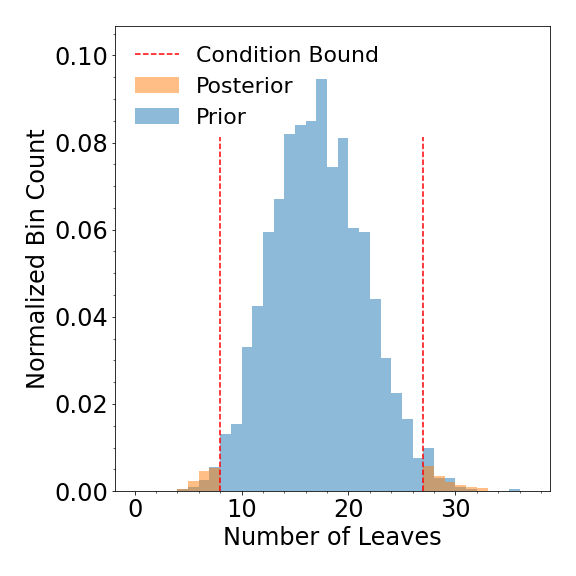}%
	}
\end{minipage}
\hspace{0.4cm}
\begin{minipage}[c]{0.45\textwidth}
\caption{\small{Histogram of the number of jet constituents (leaves) for jets generated with Ginkgo. We show the distribution of the number of constituents with no constraints (blue) and the one when using importance sampling with the limits on the number of leaves defined by the red dashed lines (orange).
}}
\label{fig:pyprob}
\end{minipage}
\vspace{-2em}
\end{figure*}

As a proof of concept, we chose to use the \texttt{PyProb} framework~\cite{baydin2019efficient}  (applied to the \texttt{Sherpa} event generator in Refs.~\cite{baydin2019etalumis}) to implement probabilistic programming in Ginkgo. After integrating \texttt{PyProb} and Ginkgo,
we successfully sampled jets while conditioning on variables, such as the jet transverse momentum and the number of constituents.
The histogram in Fig. \ref{fig:pyprob} demonstrates one simple example of the effectiveness of this framework for sampling tails of distributions.
Using \texttt{PyProb}, we are able to force Ginkgo to only produce samples of jets with fewer than eight or more than twenty-six constituents. We can see that importance sampling accurately samples the desired regions of the distribution,
i.e. with the same relative probabilities as the original distribution. 
Though this is a simple example, this method is powerful enough to allow us to condition on any sampled value within the generator, including latent variables, and condition on those values or arbitrary combinations of them. 

%% file: tex/09_discussion.tex
\section{Conclusion}
\vspace{-0.2cm}
This paper introduces a framing of common tasks in jet physics in probabilistic terms. We present \Ginkgo, a simplified generative model for jets designed to facilitate research into new computational techniques for jet physics.
A novel trellis data structure and dynamic programming algorithms that have been developed for hierarchical clustering were motivated by this area of research. The \Ginkgo\ library has been interfaced with both an implementation of the trellis and \astar algorithms and the Open AI Gym reinforcement learning library. We presented comparisons of jet clustering using greedy, beam search, Monte Carlo Tree Search, the sparse and full cluster trellis, and exact and approx. \astar search algorithms. 
These new algorithms provide a principled alternative to the generalized $k_t$ algorithms, which are based on a greedy sequential clustering algorithm. 
Additionally, we show that the trellis allows to marginalize and sample  from the true posterior distribution of clustering histories for a set of jet constituents. This could ameliorate bottlenecks when implementing the CKKW-L matching algorithm for events with large jet multiplicity. Finally, we presented how complicated regions of phase space could be sampled using probabilistic programming.

\vspace{-1em}

\section*{Acknowledgements}

We would like to thank Craig S.\ Greenberg, Nicholas Monath, Ji-Ah Lee, Patrick Flaherty, Andrew McGregor, and Andrew McCallum for their collaboration on efficient maximum likelihood estimation on Ginkgo with a trellis.
We also would like to thank Craig S.\ Greenberg, Nicholas Monath, Avinava Dubey, Patrick Flaherty, Manzil Zaheer, Amr Ahmed, and Andrew McCallum for their collaboration on exact and approximate hierarchical clustering using \astar. We are grateful for the support of the National Science Foundation under the awards ACI-1450310, OAC-1836650, and OAC-1841471, the Moore-Sloan data science environment at NYU, as well as the NYU IT High Performance Computing resources, services, and staff expertise.